\documentclass[useAMS,usenatbib]{mn2e}

\usepackage{graphicx}

\title[Ram-pressure stripping of interacting galaxies]
{On the influence of ram-pressure stripping on interacting galaxies
in clusters}

\author[W. Kapferer, T. Kronberger, C. Ferrari, T. Riser and S. Schindler]{W. Kapferer$^{1}$\thanks{E-mail:
wolfgang.e.kapferer@uibk.ac.at}, T. Kronberger$^{1}$, C.
Ferrari$^{1}$, T. Riser$^{2}$ and S. Schindler$^{1}$\\
$^{1}$Institute for Astro- and Particle Physics, University
   of Innsbruck, Technikerstr. 25, A-6020 Innsbruck, Austria\\
$^{2}$Institute for Mathematics, University
   of Innsbruck, Technikerstr. 13, A-6020 Innsbruck, Austria}

\begin{document}

\date{- -}

\pagerange{\pageref{firstpage}--\pageref{lastpage}} \pubyear{2002}

\maketitle

\label{firstpage}

\begin{abstract}
   We investigate the influence of ram pressure on the star-formation
   rate and the distribution of gas and stellar matter in interacting
   model galaxies in clusters. To simulate the baryonic and non-baryonic components of interacting
   disc galaxies moving through a hot, thin medium we use a combined
   N-body/hydrodynamic code GADGET2 with a description for
   star formation based on density thresholds. Two identical model spiral
   galaxies on a collision trajectory with three different
   configurations were investigated in detail. In the first configuration the galaxies collide without
   the presence of an ambient medium, in the second configurations the ram pressure acts
   face on on the interacting galaxies and in the third configuration the ram pressure
   acts edge on. The ambient medium is thin ($10^{-28}$ g/cm$^3$),
   hot (3 keV $\approx 3.6\times10^7$K) and has a relative
   velocity of 1000 km/s, to mimic an average low ram pressure in
   the outskirts of galaxy clusters. The interaction velocities are
   comparable to galaxy interactions in groups, falling along filaments into galaxy clusters.
   The global star formation rate of the interacting system is enhanced in the
   presence of ram pressure by a factor of three in comparison to the
   same interaction without the presence of an ambient medium. The tidal tails and
   the gaseous bridge of the interacting system are almost completely
   destroyed by the ram pressure. The amount of gas in the wake of
   the interacting system is $\sim50$\% of the total gas of the
   colliding galaxies after 500 Myr the galaxies start to feel the ram pressure.
   Nearly $\sim10-15\%$ in mass of all newly formed stars are formed in the
   wake of the interacting system at distances larger than 20 kpc
   behind the stellar discs. As the tidal tails and the gaseous bridge
   between the interacting system feel the ram pressure, knots
   of cold gas ($T<1\times10^5$K) start to form. These irregular
   structures contain several $10^6$ M$_{\sun}$ of cold gas and
   newly formed stars and, as the ram pressure acts on them, they move far away
   (several 100 kpc) from the stellar discs. They can be
   classified as 'stripped baryonic dwarf' galaxies.
   These 'stripped baryonic dwarfs' are strongly affected by turbulence,
   e.g. Kelvin-Helmholtz instabilities, which are not resolvable within the presented
   SPH simulations. Heat conduction, which is not included, would affect these small structures
   as well. Therefore we give some estimate on the lifetime of these objects.
\end{abstract}

\begin{keywords}
galaxies: interactions - intergalactic medium - galaxies: stellar
content - galaxies: structure - methods: numerical - hydrodynamics
\end{keywords}

\section{Introduction}
Multiwavelength observations have shown that the star formation rate
in interacting galaxies is enhanced in comparison to isolated
galaxies (Bushouse 1987, Sulentic 1976, Stocke 1978, Solomon \&
Sage, 1988, Combes et al. 1994, Kewley et al. 2006, and references
therein). To understand the physical processes involved in
interacting galaxies numerical simulations are an ideal tool. Since
the first publications in this field (Pfleiderer 1963, Toomre \&
Toomre 1972) it is evident that interactions are the sources of
tidal tails, bridges and other signs of massive perturbations in and
around the discs of galaxies. First calculations including star
formation and gas depletion in interacting systems (e.g. Noguchi \&
Ishibashi 1986, Noguchi 1991, Olson \& Kwan 1990 a,b, Mihos 1992,
Mihos \& Hernquist 1996) indicated that galaxy mergers are able to
increase the total star formation of the system up to an order of
magnitude and that these events of strong starbursts are able to
deplete the cold gas reservoir of the system significantly. Many
numerical investigations put special emphasis on modelling observed
interacting systems, like NGC7252 (Mihos et al. 1998) or on the
dependence of the star formation rates on interaction parameters
like spatial alignment and minimum separation (di Matteo et al.
2007, Kapferer et al. 2005). Bournaud et al. (2005) investigated the
remnants of galaxy mergers with different mass ratios; Mayer et al.
(2006) showed via numerical simulations that dwarf galaxies lose
their gas by ram pressure and tidal stripping during passages
through a disc. Cox et al. (2004) investigated galaxy mergers with
special emphasis on the heating process of gas due to shocks. The
formation of dwarf galaxies in the debris of interacting galaxies
was investigated by Duc et al. (2004). The accretion onto
supermassive black holes in merging galaxies and the resulting
suppression of star formation and the morphology of the
elliptical remnants were investigated by Springel et al. (2005).\\
From observations of the Virgo cluster of galaxies it is reported
that galaxies moving through a galaxy cluster suffer mass loss gas
by the ram pressure of the intra-cluster medium (ICM) onto the
inter-stellar medium (ISM) (Gunn \& Gott 1972, Cayatte et al. 1990,
Kenney et al. 2004, Vollmer et al. 2004). To study the effects of
ram-pressure stripping in detail several numerical approaches were
carried out, from very simple descriptions for the gas phase, like
sticky particles (e.g. Vollmer et al. 2001), to advanced Eulerian
grid techniques (e.g. Roediger et al. 2006) or Lagrangian
descriptions of hydrodynamics (e.g. Schulz \& Struck 2001,
J{\'a}chym et al. 2007), so called smoothed particle hydrodynamics
(SPH). All these methods were applied to single model disc galaxies
interacting with a time-dependent or a constant pressure from a hot
gas phase outside the disc. Recently the investigation of the
dependence of ram pressure on the star formation of a single model
galaxy moving through an ambient medium was done by Kronberger et al. (2007).\\
On larger scales the effect of ram-pressure stripping on the
chemical enrichment was studied in cosmological galaxy-cluster
simulations. To model the physics involved below the resolution of
such simulations an analytical approach for the mass loss by
ram-pressure stripping was introduced, based on the Gunn \& Gott
criterion (Schindler et al. 2005, Domainko et al. 2006). In
combination with the mass losses by galactic winds and starbursts,
ram-pressure stripping is able to enrich the ICM to observed levels
(Kapferer et al. 2007).\\
In this paper we concentrate on the distribution of baryonic,
stellar and gaseous matter, in interacting disc galaxies moving
through an ambient hot medium. By applying a density threshold
technique to model star formation (Hernquist \& Springel 2003) we
present the impact of ram-pressure stripping on the star formation
rate and the positions of star formation in comparison to galaxy
interactions without the presence of ram pressure.

\section{The simulation setup}

\subsection{The initial model for the model galaxies}
\begin{figure}
\begin{center}
\includegraphics[width=\columnwidth]{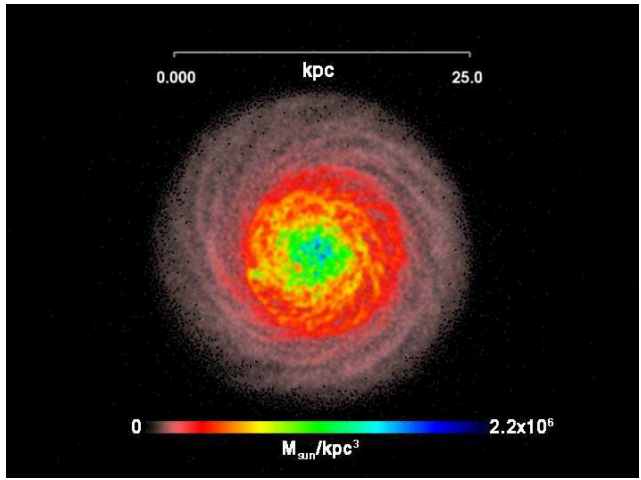}
\caption{The gas distribution in the disc of the model galaxy after
2 Gyr of evolution without ram pressure and interaction seen face
on.} \label{gal_fo}
\end{center}
\end{figure}

The model galaxies were created with an initial disc galaxy
generator developed by Volker Springel. Details and analysis can be
found in Springel et al. (2004). The total mass and the virial
radius of the galaxies' halo are given by
\begin{equation}
M_{200}=\frac{v^{3}_{200}}{10\,G\,H(z)} \qquad\mbox{and}\qquad
r_{200}=\frac{v_{200}}{10\,H(z)}
\end{equation}
with $H(z)$ the Hubble constant at redshift z and $G$ the
gravitational constant. We constructed a model galaxy with a
disc-scale length of 3.3 kpc and a circular velocity of the halo of
160 km/s. The gas fraction in the disc is initially 25\% of the
total disc mass. The mass resolution of the different components of
the galaxies (gas, stellar, dark matter) can be found in section
2.4. In Fig. \ref{gal_fo} the gas distribution of the model galaxy
after 2 Gyr of evolution without ram pressure and interaction is
shown.

\subsection{The star formation and feedback model}
We applied the so called hybrid method for star formation and
feedback introduced by Springel \& Hernquist (2003). The basic
assumption of this model is the conversion of cold clouds into stars
on a characteristic timescale $t_*$ and the release of a certain
mass fraction $\beta$ due to supernovae (SNe). The relation can be
expressed as
\begin{equation}
\frac{d\rho_{*}}{dt}=\frac{\rho_{c}}{t_{*}}-\beta\frac{\rho_{c}}{t_{*}}=(1-\beta)\frac{\rho_{c}}{t_{*}},
\end{equation}
where $\rho_*$ and $\rho_c$ are the stellar and cold gas phase
densities, respectively. The factor $\beta$ gives the mass fraction
of stars above $8 M_{\sun}$, which is in our simulations 0.1,
adopting a Salpeter initial mass function with a slope of -1.35 in
the mass interval $0.1$ to $40$ M$_{\sun}$. Each SN explosion heats
the surrounding gas in the bubble leading to an evaporation of cold
gas. The minimum temperature the gas can reach due to radiative
cooling is $10^{4}$K. From observations and analytical models it is
evident, that a certain fraction of matter can escape the galaxies
potential due to thermal and or cosmic-ray driven winds due to SN
explosions (Breitschwerdt et al. 1991). This decreases the mass and
energy budget of a galaxy, especially in the case of starbursts. To
take this into account, we applied the same method as  Springel \&
Hernquist (2003) and scale the mass outflow in such a way, that the
mass outflow is proportional to the star formation rate of the
underlying system, with a proportionally factor of two. Following
the mass budget for the hot and cold gas due to star formation, mass
feedback, cloud evaporation, growth of clouds due to radiative
cooling and mass outflow by a galactic wind lead to a selfconsistent
description of star formation in disc galaxies. More details on the
model can be found in Springel \& Hernquist (2003).

\subsection{The merger and the ram pressure}
\begin{figure}
\begin{center}
\includegraphics[width=\columnwidth]{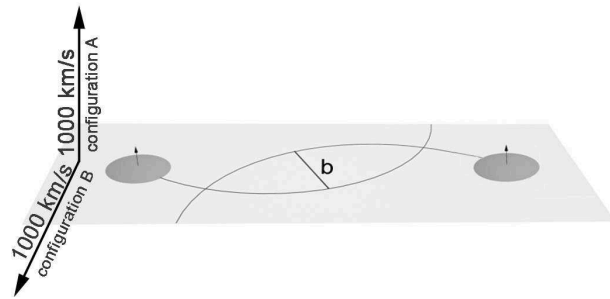}
\caption{Interaction geometry and the two configurations. The
galaxies' positions and velocities are chosen such that the galaxies
are point masses moving on Keplerian orbits. See Duc et al. (2000)
for further descriptions. In configuration A the galaxies move with
1000 km/s face on through the hot ambient medium, whereas
configuration B represents the edge on flight through the medium.
Between the configurations is an angle of 90$^{\circ}$.
Configuration C (not imprinted) is the same encounter without an
ambient medium acting on it.} \label{configurations}
\end{center}
\end{figure}

To study galaxy-galaxy interactions in a ICM wind we place the two
model galaxies on Keplerian orbits (see Duc et al. 2000 for details)
with a maximum separation of 200 kpc, a minimum separation of 1 kpc
and oriented edge on, after they have evolved for two Gyrs without
any interaction. The relative velocity of the encounter is $\sim$
300 km/s, which makes the studied merger comparable to galaxy
mergers within groups, falling along filaments into galaxy clusters.
After 2 Gyr of evolution the spiral structure of the disc galaxies
is evolved.
In addition to the interaction trajectory we gave the whole system a
directed velocity of 1000km/s through a hot ($3.6 \times 10^7$ K)
and thin ($10^{-28}$ g/cm$^3$) medium. This is done in two different
configurations. In the first configuration the galaxies fly face on
through the ICM and in the second configuration edge on while they
interact with each other, see Fig. \ref{configurations}. To compare
the effects of the ram pressure we calculate as a third
configuration of the same galaxy interaction without an ambient
medium. The galaxies collide always edge on. The simulation of the
ram pressure on the interacting galaxy pair was carried out for one
Gyr, so that the system has a first encounter but has not merged yet.\\
To study the influence of ram pressure by an ambient medium on the
distribution of the tidal tails, the gaseous bridges and the star
formation rates, we let the interaction take place in a homogenous
density and a constant temperature distribution of the ambient
medium. Therefore the effects of ram-pressure stripping are
accessible. The influence of a varying ICM temperature and density
distributions will be investigated in an upcoming work.

\subsection{The resolution}
\begin{figure}
\begin{center}
\includegraphics[width=\columnwidth]{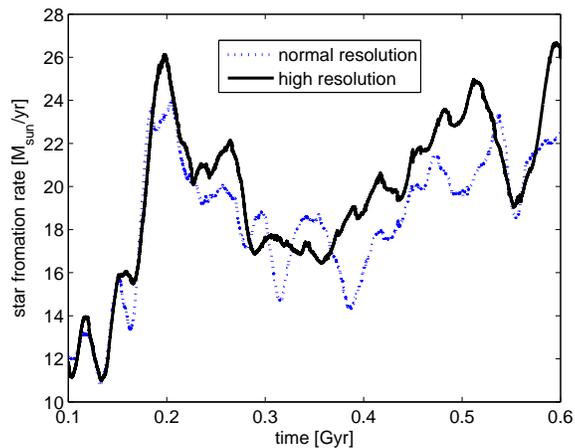}
\caption{The star formation rates for two interacting galaxies
moving edge on through the ambient medium as a function of time. The
same trend in the normal resolution and high resolution simulation
is present.} \label{sfr_hr_normal}
\end{center}
\end{figure}

The influence of the resolution on the global star formation rate of
interacting galaxies was investigated in Kapferer et al. (2005). It
was found that for interacting disc galaxies the global star
formation does not vary for mass resolutions below $10^{6}$
M$_{\sun}$ for the stellar and gas particles in the disc. To resolve
star forming regions in the bridges and tidal tails a higher
resolution was adopted in the present simulations. The total mass of
the model galaxy is $1.09 \times 10^{12}$ M$_{\sun}$, the disc mass
is $2.7 \times 10^{10}$ M$_{\sun}$ with a 25\% gas content. We
assigned $7 \times 10^{5}$ particles to each model galaxy. In Table
\ref{run properties} the particle numbers and corresponding mass
resolutions for a single model galaxy and the ICM are listed. The
ambient medium was sampled by $1 \times 10^{6}$ gas particles in a
box with 1 Mpc on a side. To study the influence of a higher
resolved ambient medium we performed a simulation with 10 times more
gas particles. The star formation rate and the distribution of the
gaseous matter did not change significantly. The same trend for the
star formation rate is present in the high resolution and the low
resolution simulation, see Fig. \ref{sfr_hr_normal}. The mean
smoothing length for ISM gas particles in our simulations is 0.978
kpc and the mean smoothing length for ICM particles is 1.17 kpc. The
gravitational softening for gas particles is 0.03 kpc, for halo
particles 0.02 kpc and for collisionless disc particles 0.05 kpc.

\begin{table}
\caption[]{Initial properties of one model galaxy and the ICM}
\begin{tabular}{c | c c c c c}
\hline \hline & number of & mass resolution & total mass\cr  &
particles & [M$_{\odot}$/particle] & [M$_{\odot}$] \cr \hline DM
halo & $3\times10^5$& $3.5\times10^6$ & $1.05\times10^{12}$\cr
gaseous disc & $2\times10^5$ & $3.4\times10^4$ & $6.8\times10^9$\cr
stellar disc & $2\times10^5$ & $1\times10^5$ & $2\times10^{10}$\cr
ICM & $1\times10^6$ & $1.46\times10^6$ & $1.5\times10^{12}$\cr ICM
hr$^1$ & $1\times10^7$ & $1.46\times10^5$ & $1.5\times10^{12}$\cr
\hline $^1$ high resolution
\end{tabular}
\label{run properties}
\end{table}

\subsection{Kelvin-Helmholtz instabilities and thermal conductivity}
Kelvin-Helmholtz instabilities develop in the case of a velocity
shear within a fluid or when there is a velocity difference along
the interface between two fluids. To resolve and treat this kind of
turbulent behaviour in hydrodynamics is a challenge. A comprehensive
analysis of the treatment of Kelvin-Helmholtz instabilities and
other turbulent phenomena, such as Rayleigh-Taylor instabilities,
was carried out with two commonly used techniques, namely Eulerian
grid-based and smoothed-particle hydrodynamics (SPH) by Agertz et
al. (2007). They conclude in their comparison that grid codes are
able to resolve and treat dynamical instabilities and mixing, while
SPH codes are not. According to Agertz et al. (2007) the reason for
this is that SPH has severe problems in the case of strong density
gradients. SPH particles representing low density regions near to
high-density regions are decoupled by erroneous pressure force
calculations due to the asymmetric density within the symmetric
smoothing kernel. This leads to a decoupling in the different phases
of the fluid, leading to an artificial suppression in the growth of
turbulent structures, such as Kelvin-Helmholtz instabilities. In
addition Agertz et al. (2007) conclude that for timescales below the
typical dynamical timescales of the turbulent structure, SPH and
Eulerian grid-based schemes agree.\\
In McCarthy et al. (2007) the stripping of a hot gaseous matter
around galaxies in groups and clusters was investigated by applying
the SPH scheme. They give an estimate for the Kelvin-Helmholtz
timescale (Eq. (4) McCarthy et al. 2007), based on the work of Mori
\& Burkert (2000) (eq.22). It is important to note that these
estimates are in principle only applicable for spherical
distributions and a particular relation between the galaxy core
radius and mass. Dropping the relation between the galaxy core
radius and mass and assuming that asphericity only introduces a
modest change, this leads to timescales in the range of Gyrs for the
main galaxies. As we stop our simulation after 1 Gyr, we conclude
that although Kelvin-Helmholtz instabilities are not well resolved
in SPH simulations our results regarding the main galaxies are not
strongly affected by it. For the small clumps the situation is
different, here the Kelvin-Helmholtz instabilities are in the range
of several hundred Myrs, assuming a relative velocity of 1000 km/s
and the low density of the ICM. Therefore the gas clumps can be
stripped completely. Many of these clumps form from gas which
originates from gas stripped into the slip stream of the galaxies.
In this situation the structures are not affected by the ram
pressure as strongly as in other regions. When entering the ICM wind
the clumps will be deaccelerated to several hundred km/s relative
velocities. This leads to longer Kelvin-Helmholtz timescales.
Another issue not addressed in the simulation is heat conduction,
which should in principle heat the gaseous clumps and therefore
prevent them from becoming denser, therefore altering the star
formation efficiency. The evaporation timescale due to thermal
conductivity of a gas clump with an n$_H$ column density of
50/cm$^3$ and a size of 1 kpc (assuming spherical geometry) in an
ambient $3.6\times10^7$K gas with an electron number density of
$10^{-1}$ is in the range of 500 Myrs, according to Nipoti \& Binney
(2007). In the case of the ICM electron number densities, which are
two orders of magnitudes smaller, these evaporation times would be
even larger. Note that the simulations presented in this work are a
simplified step towards the understanding of the complex physics
involved in environmental effects in the evolution of galaxies. The
multi-phase multi-scale gas physics, together with magnetic fields
involved in star formation, require complex and fully consistent
theoretical models which are not achieved yet.

\section{Results}

 \subsection{The distribution of the different components}
\begin{figure}
\begin{center}
\includegraphics[width=\columnwidth]{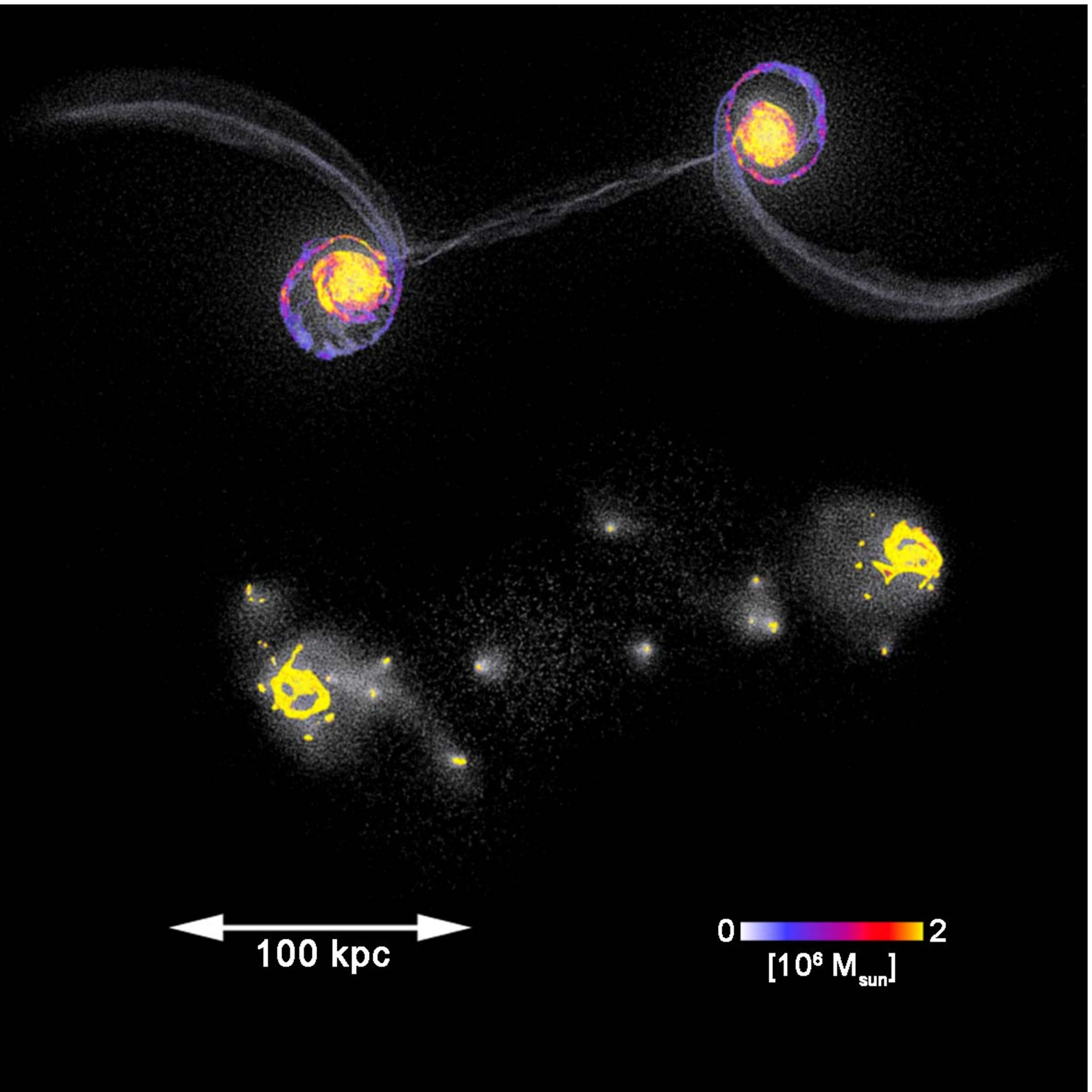}
\caption{The gas density of the interacting galaxies seen face on.
In the lower panel the interaction takes place in an ambient medium
with a constant ram pressure acting face on at the system, whereas
in the upper panel no ambient gas is present.} \label{dry_merger}
\end{center}
\end{figure}

\begin{figure}
\begin{center}
\includegraphics[width=6cm]{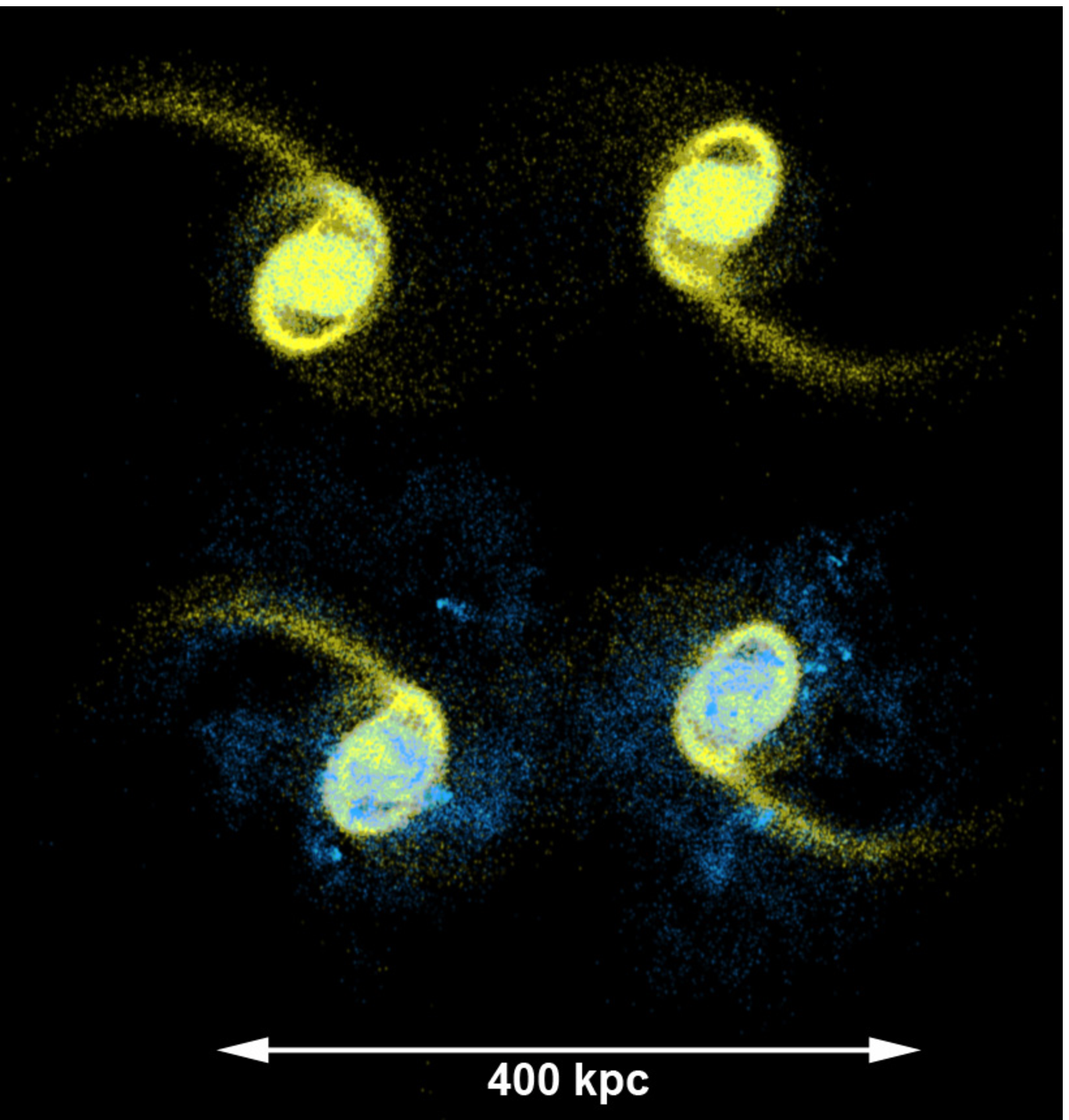}
\caption{The stellar density of the interacting galaxies seen face
on. In the lower panel the interaction takes place in an ambient
medium with a constant ram pressure acting face on at the system,
whereas in the upper panel no ambient gas is present. The new formed
stars are coloured blue, whereas the old stellar population is
coloured yellow.} \label{stars}
\end{center}
\end{figure}

\begin{figure}
\begin{center}
\includegraphics[width=\columnwidth]{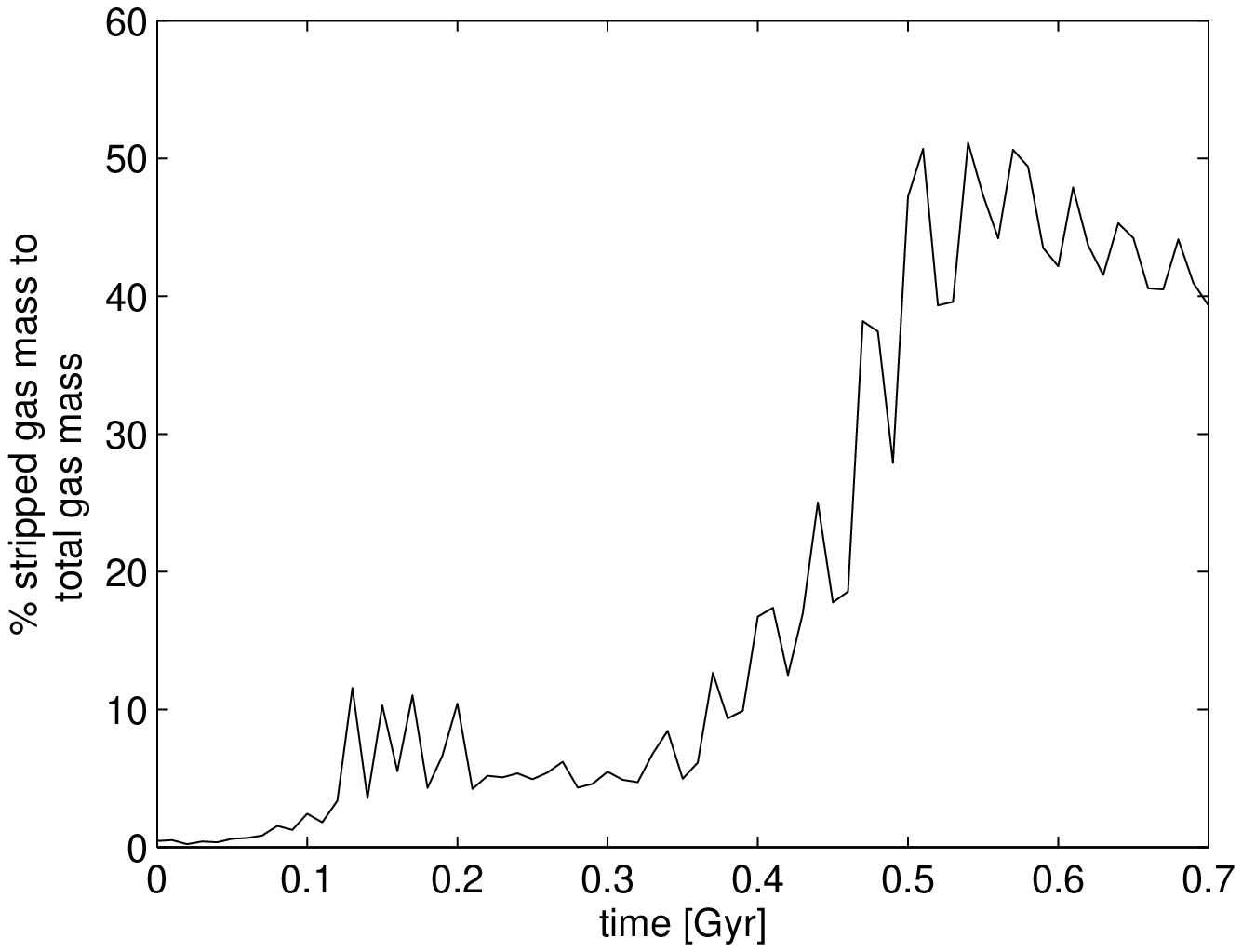}
\caption{The amount of gas in the wake (distance to the stellar disc
larger than 20 kpc) as fraction of the total amount of gas. In this
model the galaxies are moving face on through the ambient medium.}
\label{mass_wake_fo}
\end{center}
\end{figure}

\begin{figure*}
\begin{center}
\includegraphics[width=0.9 \textwidth]{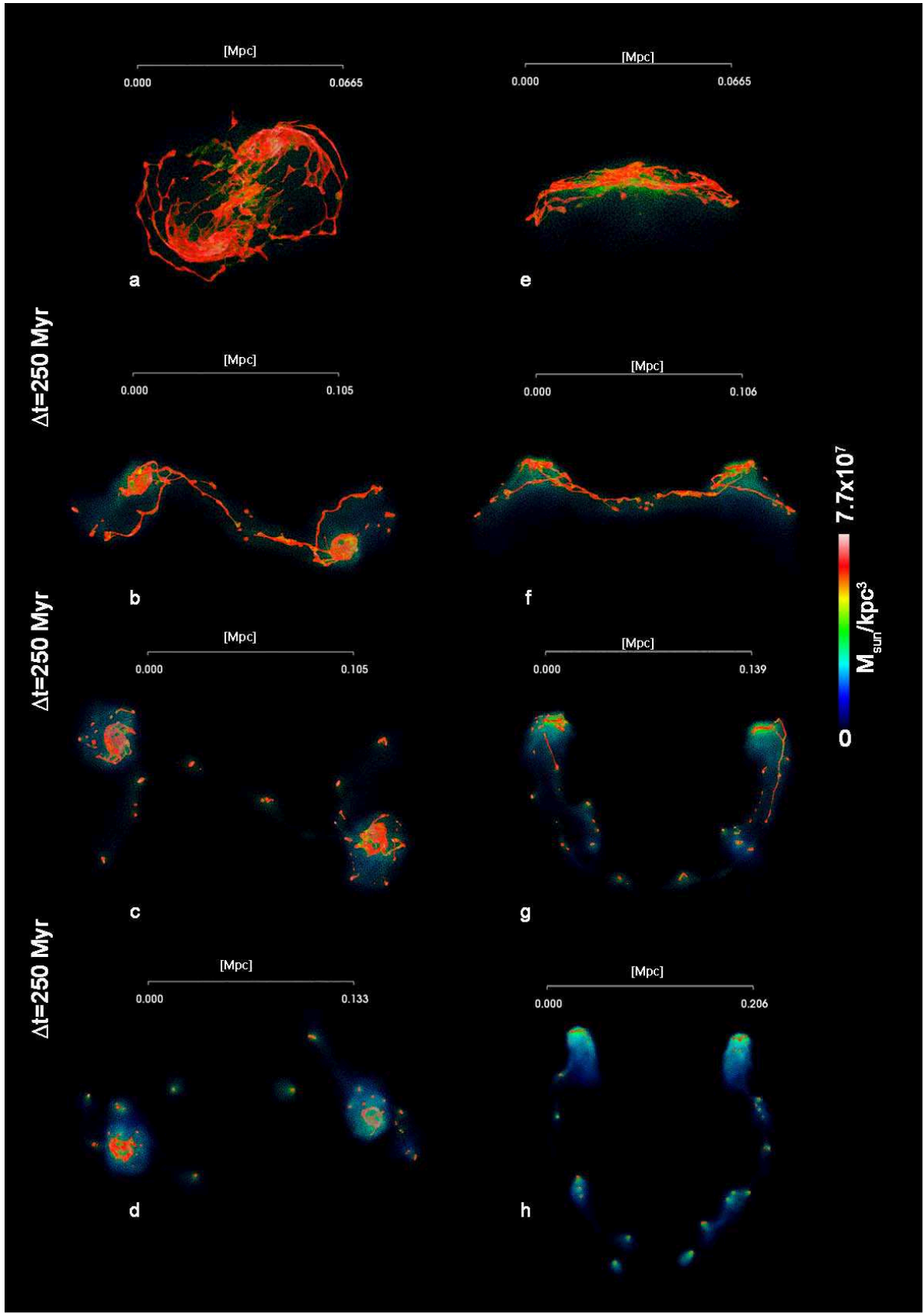}
\caption{Distribution of gas in the interacting galaxy pair exposed
to a constant ram pressure acting on the galaxies face on, seen from
different sides: left face-on, right edge-on. Between each row 250
Myr of evolution are present. The density of the ambient medium (not
shown here) has a constant density of $10^{-28}$ g/cm$^3$ and a
temperature of 3 keV. The relative velocity of the interacting pair
to the ambient medium is 1000 km/s.} \label{gas_icm_fo_time}
\end{center}
\end{figure*}

\begin{figure}
\begin{center}
\includegraphics[width=0.9 \columnwidth]{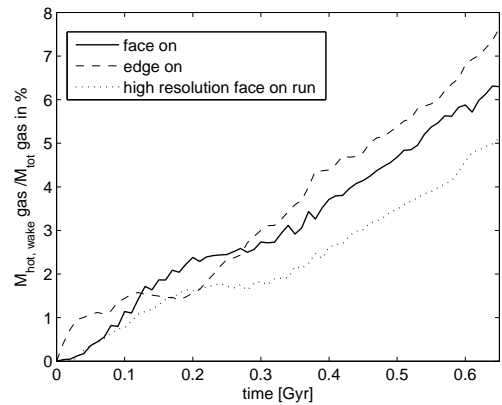}
\caption{The amount of gas originating from the interacting galaxies
in the wake (distance to the stellar disc larger than 20 kpc) and
with temperatures T$>1\times10^7$K as fraction of the total amount
of gas in the interacting system for three cases, face-on, edge-on
and for the high resolution face on simulation.}
\label{hot_mass_wake_fo}
\end{center}
\end{figure}

The most striking difference between galaxy-galaxy interactions,
suffering or not from ram pressure is the evolution of the
distribution of the gaseous tidal tails and the gas bridge between
the interacting galaxies. In Fig. \ref{dry_merger} the gas
distribution for the interacting model galaxies is shown at the
apocentre of the interaction. In the upper panel the galaxies
interact without the ram pressure acting on them, whereas in the
lower panel the effect of the ram pressure is shown. As the galaxies
move face on through the ICM the gas feels the pressure of the
ambient medium, resulting in an increasing offset of the stripped
gas, especially at the tidal tails and the bridge. The gaseous
bridge between the galaxies and parts of the tidal tails are
compressed and fragmented into massive ($\sim 10^6$ M$_{\sun}$) gas
clumps, which cool due to radiation and form new stars.
In Fig. \ref{stars} the stellar density of the interacting galaxies
seen face on is shown. In the lower panel the interaction takes
place in an ambient medium with a constant ram pressure acting face
on at the system, whereas in the upper panel no ambient gas is
present. The new formed stars are coloured blue, whereas the old
stellar population is coloured yellow.
In Fig. \ref{mass_wake_fo} the fraction of stripped gas to the total
amount of gas in the wake (distance larger than 20 kpc from the
stellar disc) is shown for an interacting system moving face on
through the ambient medium as a function of time. After 500 Myr
nearly 50\% of the gas in the interacting system is located 20 kpc
behind the stellar discs of the system. In this gaseous wake
approximately 10\% of all new stars are formed after 500 Myr.\\
The evolution of the wake is presented in Fig.
\ref{gas_icm_fo_time}. The distribution of the gas in the
interacting galaxies is shown at different timesteps. The left
column shows the face on and the right column the edge-on view. The
interval of time between each row is 250 Myr. Panel (a) and (e) show
the situation at the first encounter. The ram pressure already
distorts the outer gas layers. The effect of the ram pressure on the
tidal tails and bridges starts after the galaxies have had the first
encounter. In panels (b) and (f) the compression of the gas is
already distinct. As the gas gets more compressed it cools and
becomes denser. Parts of the cooled gas fall back onto the discs
leading to episodes of star formation in the disc. In panels (c) and
(g) the tidal tails and the gaseous bridge are nearly completely
destroyed. Only dense knots are visible around the discs at
distances larger than 50 kpc. The knots are star forming regions
with several $10^5$ M$_{\sun}$ of newly formed stars. The last
timestep shows that the dense knots of gas become more and more
separated from the interacting system. While these knots form stars,
more and more gas is transformed into stellar matter, leading to
small irregular structures with several $10^6$ M$_{\sun}$ of baryonic matter.\\
The evolution of the ratio of heated gas (T$>10^7$K) in the wake
originating from the interacting system to the total gas of the
system is given in Fig. \ref{hot_mass_wake_fo}. After 0.7 Gyr
$\sim$5-6\% of the stripped material in the interacting system is
heated to temperatures above T$>10^7$K. This gas will show its
signatures in X-ray observations of the system.

\begin{figure}
\begin{center}
\includegraphics[width=\columnwidth]{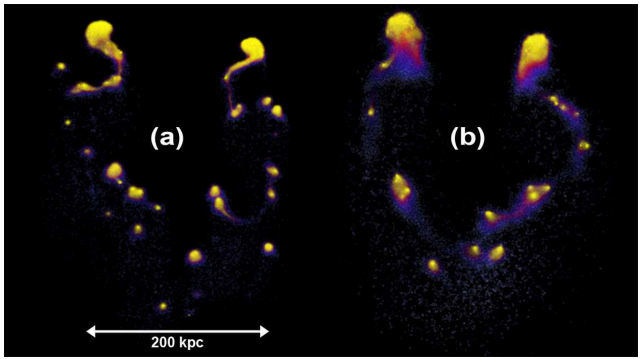}
\caption{The distribution of gas in the interacting galaxies after
one Gyr of evolution in the high resolution ICM simulation (a) and
normal resolution ICM simulation (b), seen edge on.}
\label{hr_lr_density}
\end{center}
\end{figure}

In the case of the galaxies moving edge on through the ambient
medium the distribution of matter is very similar. The gaseous
bridge and the tidal tails are destroyed by the ram pressure. Dense
knots of gas form, which cool in the same way as in the face on
situation. The remaining gaseous discs are truncated very similarly
as in the face on passage.\\
We have additionally investigated the robustness of our results with
respect to the resolution. In Fig. \ref{hr_lr_density} the
distribution of gas of the interacting galaxies after one Gyr of
evolution in the high resolution ICM simulation (a) and normal
resolution ICM simulation (b) are shown. In the highly resolved ICM
simulation the knots of gas, which form from stripped matter in the
gaseous bridges and the tidal tails are present as in the normal
resolved ICM simulation. The calculated amount of gas in the wake,
originating from the interacting galaxies, defined as gas which
lacks 20 kpc behind the discs, is 48\% in the case of the highly
resolved ICM simulation and 43\% in the normal resolved simulation.
As the ram-pressure is the dominating mechanism, galactic outflows
do not alter the mass distribution significantly. The amount of
gaseous and stellar matter in the wake of the interacting galaxies
is not changed.

\subsection{The star formation rate and the regions of star formation}

\begin{figure}
\begin{center}
\includegraphics[width=\columnwidth]{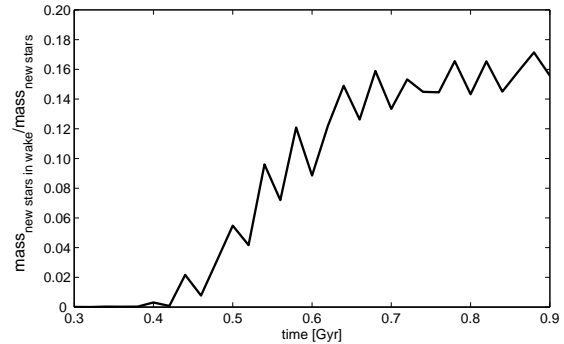}
\caption{The evolution of the ratio of newly formed stars in the
wake to the total amount of newly formed stars in the case of the
face on interaction. The definition for the stars in the wake is a
distance larger than 20 kpc to the interacting discs.}
\label{stars_wake}
\end{center}
\end{figure}

\begin{figure}
\begin{center}
\includegraphics[width=\columnwidth]{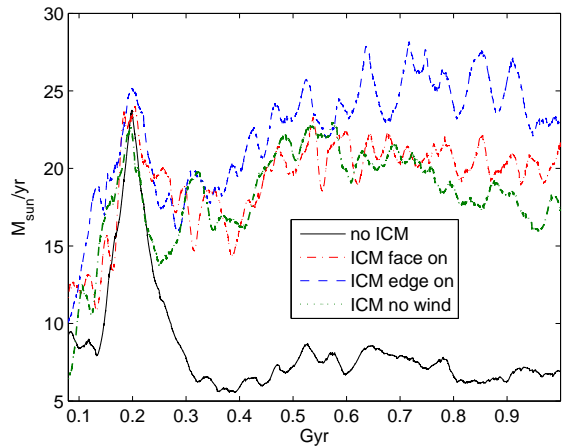}
\caption{The star formation rate of the the merging system as a
function of time for four different configurations. The black solid
line shows the star formation rate for the interacting system
without ambient medium, the dashed blue line gives the face on case
of the ram pressure, the dotted red line the edge-on ram-pressure
case, corresponding to configurations A and B in Fig. 1. The dotted
green line gives the star formation rate for an interaction taking
place in the ambient medium without a constant wind.} \label{sfrs}
\end{center}
\end{figure}

\begin{figure*}
\begin{center}
\includegraphics[width=0.9 \textwidth]{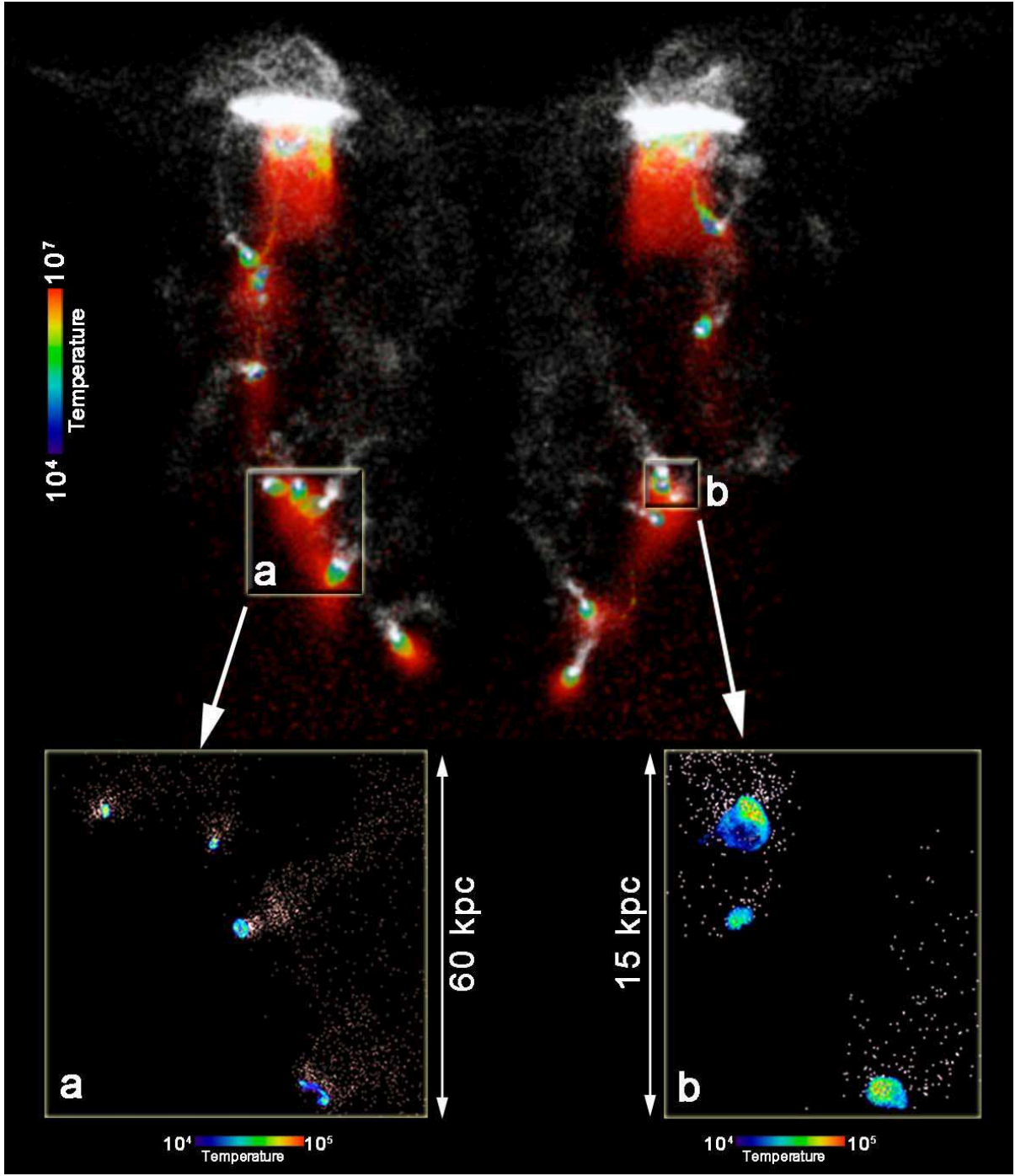}
\caption{The gas and stellar distribution of matter in the wake of
two interacting galaxies moving face on through an ambient thin
($10^{-28}$ g/cm$^3$), hot ($5\times10^7$ K) medium with 1000 km/s.
The gaseous component of the wake is colour coded for the
temperature, whereas the newly formed stars in the system are shown
in white colour. The inserts (a) and (b) give two irregular
structures in the wake of the interacting galaxies in more detail.
Note that only the cold gas and newly formed stars are shown in the
inserts. The total gas mass in insert (a) is $3.5\times10^6$
M$_{\sun}$, the total stellar mass present in the volume shown by
insert (a) is $3.4\times10^6$ M$_{\sun}$. The total gas mass located
in the region shown in insert (b) is $2.3\times10^6$ M$_{\sun}$ and
the total stellar mass present in insert (b) is $5.9\times10^5$
M$_{\sun}$.} \label{zoo_fo}
\end{center}
\end{figure*}

To study the effects of constant ram pressure on the star formation
we did as a first step simulations with a single disc galaxy moving
through a hot medium with 1000 km/s relative velocity. The results
of these investigations are discussed in detail in Kronberger et al.
(2008). In the case of the interacting galaxies in an ambient
medium, newly formed stars can be found up to a hundred kpc behind
the plane of the disc. In Fig. \ref{stars_wake} the evolution of the
ratio of newly formed stars in the wake to the total amount of newly
formed stars in the case of the face-on interaction is shown. The
definition for newly formed stars in the wake is a distance larger
than 20 kpc to the interacting discs. Almost 20 \% of all newly
formed stars are located at a distance of more than 20 kpc to the
interacting discs after a simulation time of roughly one Gyr.
The evolution of the total star formation rate for the interacting
systems is shown in Fig. \ref{sfrs}. In the case of the interaction
without an ambient medium the first encounter increases the star
formation rate by a factor of $\sim3$. After the first encounter the
global star formation rate decreases to slightly smaller rates than
before. This can be explained by the formation of the tidal tails
and the bridge, which decrease the gas content of the discs.\\
In the case of an external ram pressure the situation is very
different. After the first encounter the star-formation enhancement
does not decrease. This different behaviour is present in both
configurations, face on and edge on acting ram pressure. The reason
for the enhancement can be found in the compression and deformation
of the tidal tails and the gaseous bridges, as well as the
compression of the discs due to the external pressure of the ICM on
it. By comparing the star formation rates with a simulation in the
ICM with no directed wind reveals, that most of the enhancement,
belongs to the pressure of the ICM onto the galaxies. As the bridges
and the tails are affected in the same way in the case of face on
and edge on ram pressure, the total star formation rate does not
vary significantly between the two configurations. The influence of
the outflow of a galactic wind on the star formation rate is minor.
As the outflow is kept in the range of twice the star formation
rate, which is in agreement with observations (Martin, 1999), it is
a negligible mass loss of the interacting system in comparison to
the amount of stripped matter. The cool, dense knots and the
distorted gas in the discs form new stars.
In Fig. \ref{zoo_fo} the star-forming regions in the simulation of
face on ram pressure are shown. The upper panel shows an edge-on
view of the interacting system with the temperature colour-coded gas
distribution. Additionally the newly formed stars are highlighted as
white points. The lower panels give two star forming regions (a) and
(b) in a larger view. Only the cool gas (T$<10^5$ K) and newly
formed stars are shown. The total gas mass, originating from the
interacting system in insert (a) is $3.5\times10^6$ M$_{\sun}$, the
total stellar mass present in the volume shown by insert (a) is
$3.5\times10^6$ M$_{\sun}$. The total gas mass, stripped from the
interacting system and presented in insert (b) is $2.3\times10^6$
M$_{\sun}$ and the total stellar mass present in insert (b) is
$5.9\times10^5$ M$_{\sun}$.

\section{Comparison to observations}

As the merger investigated in this work is more likely happening in
groups, located in the infall regions of galaxy cluster, a direct
comparison to observations of mergers within galaxy clusters is
restricted. Nevertheless some observations, especially those in the
outskirts of galaxy clusters, can be compared to study the influence
of a low ram pressure. Since the first HI radio atlas of the Virgo
cluster by Cayatte et al. (1990), it was found that ram-pressure
stripping affects the star formation of cluster galaxies. Up to now,
X-ray observations revealed the existence of tails or wakes of gas,
probably associated to ram-pressure stripping, in about ten
elliptical cluster galaxies or group galaxies (Kim et al. 2007,
Jaltema et al. 2008 and Schindler \& Diaferio 2008 and references
therein). Only in recent years optical, radio and X-ray observations
have started to accumulate evidence of gas tails stripped from
late-type cluster galaxies at hundred kpc scales (e.g. UGC 6697,
CGCG 97-073/97-079 in A1367: Gavazzi et al. 2001a,b, Sun \&
Vikhlinin 2005 -- C153 in A2125: Wang et al. 2004 -- ESO 137-001 in
A3627: Sun et al. 2006, 2007 -- NGC 4525, NGC 4388 and NGC 4532/DDO
137 in Virgo: Yoshida et al. 2002, Oosterloo \& van Gorkom 2005,
Haynes et al. 2007, Koopmann 2007 -- D100 in Coma: Yagi et al.
2007). Most of these galaxies show signatures of enhanced star
formation. The fact that some of them are interacting objects is
well proven (e.g. CGCG 97-073/97-079 in A1367: Gavazzi et al. 2001b,
NGC 4532/DDO 137 in Virgo, Koopmann 2007), still debated in other
cases (e.g. UGC 6697 in A1367, Gavazzi et al. 2001a, NGC 4254 in
Virgo, Vollmer et al. 2005). Concerning this point, it is worth to
be stressed here that ram-pressure deeply alters the
``expected''\footnote{Based on
  the results of previous models of galaxy interactions, which do not
  take into account the ambient medium effects (e.g. Kapferer et
  al. 2005).} morphology of interacting galaxies, as clearly shown in
Fig. \ref{dry_merger}. This point is extremely important to properly
identify the different physical mechanisms acting on galaxies when
interpreting observational results.\\
Since stripped tails are diffuse sources that require high
sensitivity observations with a sufficiently large field of view,
only for very few of them we have detailed information coming from
multi-wavelength data. Our simulations indicate that we should be
able to detect in X-rays heated gas (T $> 10^7$ K) in the wake of
interacting systems. At least three extended X-ray tails of
late-type galaxies have been observed in clusters (UGC 6697 in
A1367: Sun \& Vikhlinin 2005 -- ESO 137-001 in A3627: Sun et al.
2006 -- C153 in A2125: Wang et al. 2004). The gas temperature
measured in two of these sources is in the expected range (T $\sim
0.5 - 1.2 \times 10^7$ K, Wang et al. 2004, Sun \& Vikhlinin 2005).
In ESO 137-001 the X-ray tail coincides positionally with a detected
H$\alpha$ tail and 29 HII regions have been detected in the wake of
gas (Sun et al. 2007). Similarly to the results of our simulations,
the HII regions closest to the galactic disc form a bow-like front
with the axis nearly in the same direction as the tail.
Additionally, Sun et al. (2006) detected three possible
ultra-luminous X-ray sources (ULXs) related to active star formation
in the tail. In UGC 6697 hints of correlation between X-ray and
H$\alpha$ emission have also been shown (even if on smaller scales).
Additionally, at least two X-ray point sources have been detected in
the tail of this galaxy, which may be associated with star clusters.
The detailed analysis of Sun \& Vikhlinin (2005) also demonstrates
that ram pressure alone cannot explain the peculiarities of UGC 6697
(i.e. its complex velocity field, the presence of a warp in the
South-East region of the stellar disk). In agreement with our
numerical results, they thus suggest that, together with ram
pressure stripping, tidal effects related to galaxy interactions
could play a role in determining the observed properties of UGC 6697
and its X-ray tail.\\
On the other hand, it has also been proven that tidal effects alone
cannot explain the observational properties of some interacting
galaxies. Two irregular objects (CGCG 97-073 and 97-079) located in
the North-West region of the galaxy cluster A1367 show extended
($\sim$75 kpc) tails of H$\alpha$ and synchrotron emission (Gavazzi
et al. 1995, 2001b). Observations indicate that the tails host
$\sim$40\% of the original gas of the two galaxies. Gavazzi et al.
(2001b) concluded that both the high SFR of these irregular galaxies
and their head-tail morphology can be explained by ram-pressure
effects. However, due to the low ICM density in the peripheral
cluster region where the two galaxies are located, they claimed the
need of an additional physical mechanism (i.e. the interaction
between the two objects), able to lose their potential well, thus
making ram-pressure more effective in stripping and/or compressing
the ISM.\\
The fraction of stripped gas in the wake of a given galaxy is
significantly higher (and comparable to the observed value) in the
case of interacting systems than in isolated ones (see Kronberger et
al. 2008). The combination of a galaxy encounter and of moderate
ram-pressure was already suggested to be responsible for the
perturbed atomic gas distribution observed in interacting galaxies
(NGC 4654/NGC 4639 and NGC 4254 in the Virgo cluster, Vollmer 2003,
Vollmer et al. 2005). Haynes et al. (2007) and Duc et al. (2007)
have recently suggested that more extended ($\geq$ 100 kpc) HI tails
can be created by galaxy harassment and high-speed galaxy
collisions. Our simulations show that wakes of gas and newly formed
stars of several hundred of kpc are originated in interacting
galaxies subject to ram pressure stripping. In addition, our results
can explain: a) the origin of the blue colour and of the ionized gas
detected in several gas wakes of interacting/disturbed galaxies
(e.g. Gavazzi et al. 2001a, Yoshida et al. 2002, Sun \& Vikhlinin
2005, Sun et al. 2007), and b) the existence of ULXs and isolated
intra-cluster HII regions, observed at several tens of kpc from the
closest galaxy and related to stars that have formed within the
intracluster volume (e.g. Gerhard et al. 2002, Sun et al. 2007).\\
Finally, our results can explain the origin of the region with the
highest density of star formation activity ever observed in a local
cluster (A1367, Cortese et al. 2006). Two giant galaxies and several
dwarfs/extragalactic HII regions, together with an extended
($\geq$150 kpc) wake of ionized gas, were observed in a compact
group infalling towards the cluster centre. Both ram-pressure
stripping and galaxy-galaxy interactions are extremely efficient in
this case. Due to their lower velocity dispersion compared to
clusters, galaxy groups are actually the natural site for tidal
interactions. This group is furthermore not isolated, but moving
with a high velocity ($\sim$1700 km/s) in the ICM of A1367 (Cortese
et al. 2004). Note that the simulation presented in this work is not
representative for high velocity encounters/mergers in the central
regions of galaxy clusters. Here additional effects, like galaxy
harassment, play an important role. On the other hand the ram
pressure increases with the ICM density and the square of the
relative velocity of the galaxy with respect to the ICM. Therefore
the effect of ram pressure stripping increases in the denser regions
of galaxy clusters. The relative strength and the interplay of ram
pressure in high velocity encounters will be investigated in an
upcoming paper.

\section{Discussion and conclusions}
In order to investigate the influence of the ram pressure on
interacting galaxies, we compared the results of a simulation of an
interaction with and without a constant wind. We focus on the
evolution of the star formation rate of the interacting system and
on the distribution of the different baryonic components (i.e. gas
and stars). The results can be summarised as follows:

\begin{itemize}
\item The star formation rate of the interaction is enhanced in the
presence of an ambient hot (3keV) and rare medium ($10^{-28}
$g/cm$^3$) together with ram pressure by a factor of three in
comparison to the same interaction without the ambient medium.

\item The morphology of the interacting system is strongly distorted.
The tidal tails and the bridge between the interacting system are
completely destroyed by the ram pressure. The resulting gas and
stellar mass distributions of the two galaxies would not be
characterised by observers as interacting system after the first
close encounter.

\item The amount of gas in the wake of the interacting system
is 50\% of the total gas mass of the interacting system after 500
Myr of ram pressure acting on it. Approximately 10-15\% of the gas
is heated up to temperatures above $1\times10^7$K, which would be
observable in X-rays.

\item After 500 Myr of ram pressure $\sim10\%$ of all newly formed
stars are formed in the wake of the interacting system at distances
larger than 20 kpc behind the stellar disc in the case of the face
on ram pressure. The same behaviour can be observed in the case of
the edge on ram pressure.

\item As the tidal tails and the gaseous bridge between the
interacting system feel the ram pressure, knots of cold gas
($T<10^5$K) start to form and these irregular structures start to
form new stars. These knots are found to contain several $10^6$
M$_{\sun}$ of cold gas and newly formed stars. They can be
classified as 'stripped baryonic dwarf' galaxies. The lifetime of
these 'stripped baryonic dwarfs' is limited by turbulence and heat
conduction. If the objects are in the slipstream of the disc
galaxies, they can survive for a several hundred Myrs up to a Gyr.
As the ram pressure in the gaseous wake is decreasing due to the
decreasing relative velocities, the timescales for Kelvin-Helmholtz
instabilities is increasing as well, therefore extending the
lifetime of these objects. Thermal conduction is affecting 'stripped
baryonic dwarfs' as well, but the low densities of the surrounding
ICM seems to result in evaporation timescales at least larger than
500 Myrs.
\end{itemize}

\noindent Concluding we found that interacting galaxies affected by
a moderate ram pressure show a completely different behaviour
compared to the case of no ram pressure acting on them. The
simulations presented in this work are idealised in order to be able
to distinguish between different effects. As galaxies move through a
real cluster, the ram pressure changes and complex interactions with
the cluster potential and other galaxies are present. These factors
will be investigated in upcoming papers.

\section*{Acknowledgements} We thank the anonymous referee for fruitful comments
which helped to improve the quality of the paper. The authors thank
Volker Springel for providing them with GADGET2 and his
initial-conditions generator. The authors acknowledge the Austrian
Science Foundation (FWF) through grants P18523-N16 and P19300-N16.
Thomas Kronberger is a recipient of a DOC fellowship of the Austrian
Academy of Sciences. The authors further acknowledge the
UniInfrastrukturprogramm des BMWF Forschungsprojekt Konsortium
Hochleistungsrechnen, the ESO Mobilit\"atsstipendien des BMWF
(Austria), and the Tiroler Wissenschaftsfonds (Gef\"ordert aus
Mitteln des vom Land Tirol eingerichteten Wissenschaftsfonds).

\end{document}